\documentclass{PoS}

\usepackage{amssymb}
\usepackage{epsfig,graphicx}

\title{ChPT tests at the NA48 and NA62 experiments at CERN}

\ShortTitle{ChPT tests at the NA48 and NA62 experiments at CERN}

\author{\speaker{Dmitry MADIGOZHIN}%
         \thanks{for the NA48/2 and NA62 Collaborations:
F.~Ambrosino, A.~Antonelli, G.~Anzivino, R.~Arcidiacono,
W.~Baldini, S.~Balev, J.R.~Batley, M.~Behler, S.~Bifani, C.~Biino
% ...
}\\
        ( JINR, Dubna)\\
        E-mail: \email{Dmitri.Madigojine@cern.ch}
}

\abstract{
The NA48/2 Collaboration at CERN has accumulated unprecedented statistics of rare kaon decays in the $K_{e4}$ modes: 
$K_{e4}(+-)$ ($K^\pm \to \pi^+ \pi^- e^\pm \nu$) and $K_{e4}(00)$ ($K^\pm \to \pi^0 \pi^0 e^\pm \nu$) with nearly one 
percent background contamination. 
The detailed study of form factors and branching rates, based on these data, has been completed recently. The results 
brings new inputs to low energy strong interactions description and tests of Chiral Perturbation Theory (ChPT) and 
lattice QCD calculations. In particular, new data support the ChPT prediction for a cusp in the $\pi^0\pi^0$ invariant 
mass spectrum  at the two charged pions threshold for $K_{e4}(00)$ decay. 
New final results from an analysis of about 400 $K^\pm \to \pi^\pm \gamma \gamma$ rare decay candidates 
collected by the NA48/2 and NA62 experiments at CERN during low intensity runs with minimum bias trigger configurations 
are presented. The results include a model-independent decay rate measurement and fits to ChPT description.
}

\FullConference{XIIth International Conference on Heavy Quarks \& Leptons 2014\\
		 25-29 August 2014\\
		 Schloss Waldthausen, Mainz, Germany}

\begin{document}

\section{Introduction}
\label{intro}

Investigation of rare kaon decays allows for the determination of Chiral Perturbation Theory (ChPT) 
constants and provides the basis for its validity checks. Analysis of high statistics data samples
collected by NA48/2 and NA62($R_K$ phase) experiments in several decay modes allows the stringent
tests of ChPT predictions.  

The main goal of NA48/2 experiment was the search for CP-violating asymmetry in $K^\pm \to 3\pi^\pm$ 
decays \cite{Batley:2007}. Additionally, it has provided in 2003-2004 a large data sample for charged 
kaon rare decay studies. In 2007-2008, the NA62 experiment \cite{Anelli:2005ju} ($R_K$ phase) has collected 
another large data sample with the same detector (described in \cite{Fanti:2007vi}) but modified beam line.

\begin{figure}
\begin{center}
\setlength{\unitlength}{1mm}
\resizebox{\columnwidth}{!}{%
\begin{picture}(100.,40.)                
\includegraphics[width=100mm]{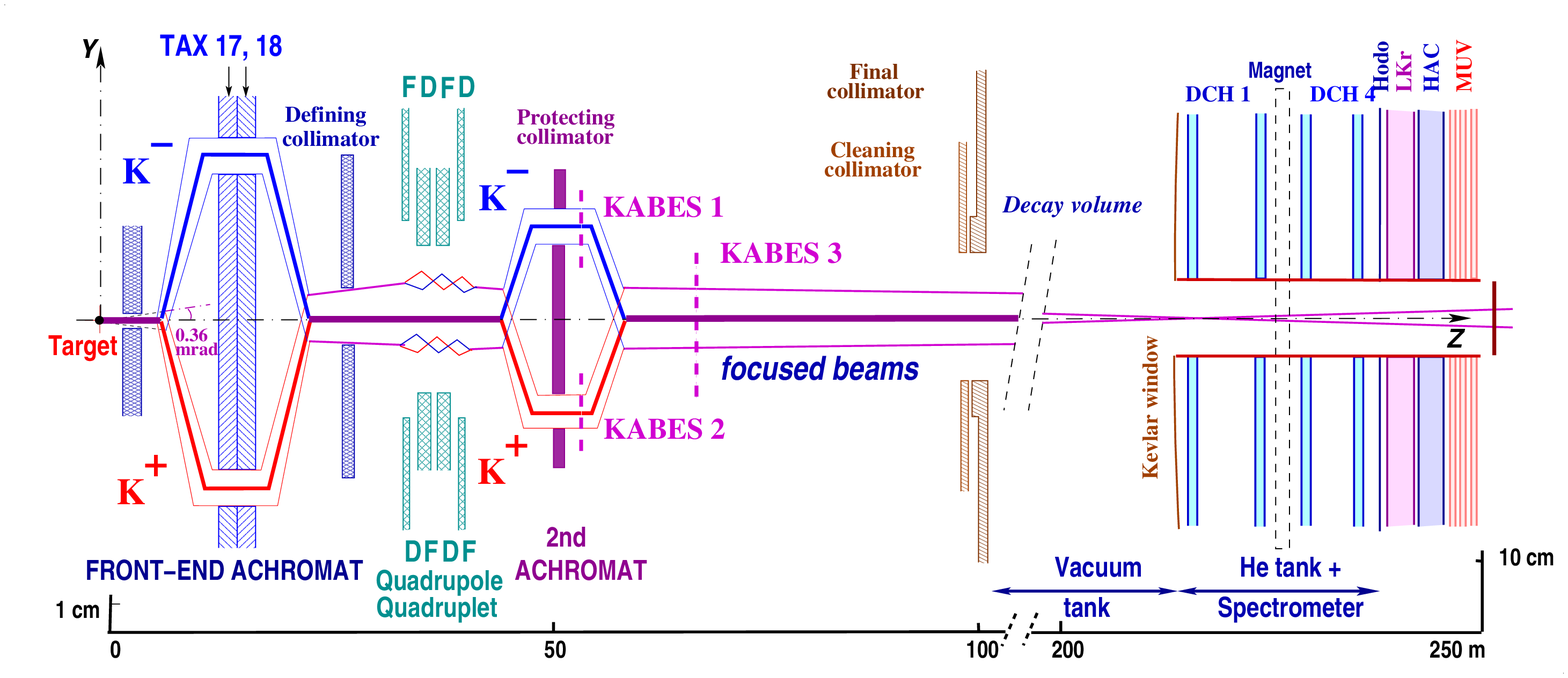}
\end{picture}
}
\end{center}
\caption{NA48/2 beamline}
\label{beamline}
\end{figure}

Two simultaneous $K^+$ and $K^-$ beams were produced by $400$~GeV/$c$ ~protons 
on a beryllium target (Fig. \ref{beamline}). Particles of opposite charge with a central momentum of 
$60$~GeV/$c$ and a momentum band of $\pm 3.8\%$ ($rms$) ($74$~GeV/$c$ $\pm 1.9\%$ for NA62 $R_K$ phase) 
were selected by the system of magnets and collimators.
Both beams of about 1 cm width were following almost the same path in the decay volume contained in 
a 114 m long vacuum tank. The beams were dominated by $\pi^\pm$, the kaon component was about 6\%.

Charged products of $K^\pm$ decays were measured by the 
magnetic spectrometer consisting of four drift chambers (DCH1--DCH4) and a dipole magnet located between DCH2 and DCH3. 
The spatial resolution of each DCH was nearly $90 \mu m$ and the momentum resolution was 
$\frac{\sigma_p}{p} = (1.02 \oplus 0.044 \cdot p)\%$ ($p$ in Gev/$c$). 
The spectrometer was followed by a scintillator hodoscope with a time resolution of \~ 150 ps, 
whose fast signals were used to trigger the readout of events with a charged track. 

A Liquid Krypton calorimeter (LKr), located behind the hodoscope, was used to measure the energy of 
electrons and photons. It is an almost homogeneous ionization chamber with an active volume of $7\ m^3$ of 
liquid krypton 27 $X_0$ deep, segmented transversally into projective cells, $2 \times 2\ cm^2$ each. 
Transverse position of isolated shower was measured with a spatial resolution 
$\sigma_x = \sigma_y = (0.42/\sqrt{E} \oplus 0.06)\ cm$. Energy resolution for photons and electrons was
$\sigma_E/E = (3.2/\sqrt{E} \oplus 9.0/E \oplus 0.42)\%$ (E in GeV).

An aluminium beam pipe of 16 cm outer diameter and 1.1 mm thickness was traversing the centres of all the detector 
elements, providing the path in vacuum for undecayed beam particles and for muons from beam $\pi^\pm$ decays.

\section{$K^{+-}_{e4}$ decay}
\label{kpme4}

\begin{figure}
\begin{center}
\setlength{\unitlength}{1mm}
\resizebox{0.70\columnwidth}{!}{%
\begin{picture}(100.,40.)                
\includegraphics[width=100mm]{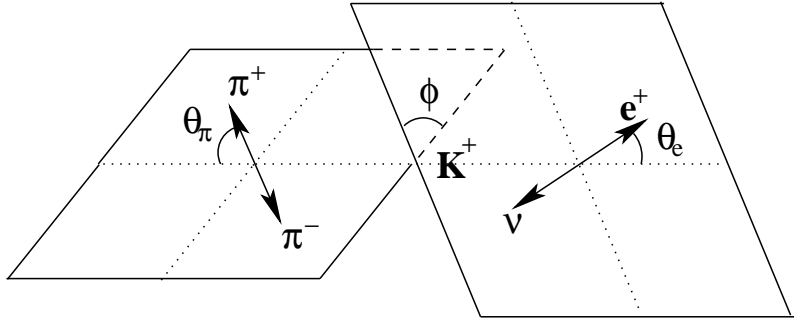}
\end{picture}
}
\end{center}
\caption{Topology of $K_{e4}(+-)$ decays.}
\label{topology}
\end{figure}

Kinematics of the $K^{\pm} \rightarrow \pi^+ \pi^- e^{\pm} \nu$ ($K^{+-}_{e4}(+-)$) 
decay is defined by five variables \cite{Cabibbo:1965zz}:
the squared invariant masses of dipion ($S_{\pi}$) and dilepton ($S_e$), the angle 
$\theta_{\pi}$ of $\pi^{\pm}$ in dipion rest frame with respect to the flight direction 
of dipion in the kaon center of mass system,  the similar angle $\theta_e$ of $e^{\pm}$ in 
dilepton rest frame, and the angle $\phi$ between dipion and dilepton planes (see Fig. \ref{topology}).

$K_{e4}$ decay amplitude is a product of the leptonic weak current and (V-A)
hadronic current, described in terms of three (F,G,R) axial-vector and one (H) 
vector complex form factors. Sensitivity of $K_{e4}$ decay matrix element to R form 
factor is negligible due to the small mass of electron.
Form-factors may be developed in a partial wave expansion: 
\begin{eqnarray}
\nonumber
F = F_s e^{i \delta_{fs}} +  F_p e^{i \delta_{fp}}  cos \theta_{\pi} + F_d e^{i \delta_{fd}}  cos^2 \theta_{\pi} + ...\\
\nonumber
G = G_p e^{i \delta_{gp}} +  G_d e^{i \delta_{gd}}  cos \theta_{\pi} + ...\\
H = H_p e^{i \delta_{hp}} +  H_d e^{i \delta_{hd}}  cos \theta_{\pi} + ...
\end{eqnarray}

Limiting the expansion to S- and P-waves and considering a unique phase $\delta_p$ for all P-wave
form factors in absence of CP violating weak phases, one will obtain the decay probability, that 
depends only on the real form factor magnitudes $F_s,F_p,G_p,H_p$, a single phase shift 
$\delta=\delta_s-\delta_p$ and kinematic variables.
The partial wave form factors can be developed in a series expansion of the dimensionless invariants 
$q^2 = (S_\pi /4m^2_{\pi})-1$ and $S_e/4m^2_{\pi}$ \cite{Amoros:1999mg}.
Two slope and one curvature terms are sufficient to describe the $F_s$ form factor variation 
within the available statistics ($F_s/f_s = 1+(f'_s/f_s) q^2 + (f''_s/f_s) q^4 + (f'_e/f_s) S_e/4m^2_{\pi}$),
while two terms are enough to describe the $G_p$ form factor ($G_p/f_s = g_p/f_s + (g'_p/f_s) q^2$),
and two constants -- to describe the $F_p$ and $H_p$ form factors. 

Hadronic form factors for the S- and P-waves have been obtained 
by NA48/2 concurrently with the phase difference between the S- and P-wave states of $\pi \pi$ 
system, leading to the precise determination of $a^0_0$ and $a^0_2$, the I=0 and I=2 S-wave 
$\pi \pi$ scattering lengths \cite{Batley:2009zz}. 

A high precision measurement of $K^{+-}_{e4}$ form factors and branching fraction
has been published by NA48/2 few years later \cite{Batley:2010zza,Batley:2012rf}. 
$K_{e4}^{+-}$ decay rate was measured relative to $K^{\pm} \to \pi^+\pi^-\pi^{\pm}$ ($K_{3\pi}^{+-}$) 
normalization channel. 

For the selection in this analysis a track of charged particle with a momentum 
$p>2.75$~GeV and $0.9 < E/p < 1.1$ was identified as $e^{\pm}$, while the track with 
$p>5$~GeV and $E/p < 0.8$ was regarded as $\pi^{\pm}$.
A dedicated linear discriminant variable based on shower properties has 
been applied to reject events with one misidentified pion. To suppress $K_{3\pi}^{+-}$ background, the 
vertex invariant mass $M_{3\pi}$ in the $\pi^+\pi^-\pi^{\pm}$ hypothesis and its transverse momentum 
$p_t$ were required to be outside an ellipse centered at PDG kaon mass \cite{Beringer:1900zz} 
and zero transversal momentum, with semi-axes of $20\ MeV/c^2$ and $35\ MeV/c$, respectively. 

The squared missing mass was required to be $ > 0.04\ (Gev/c^2)^2$ to reject 
$\pi^{\pm}\pi^0$ decays with a subsequent $\pi^0 \to e^+e^-\gamma$ process. 
The invariant mass of $e^+e^-$ system was required to be $> 0.03\ GeV/c^2$ in order 
to reject photon conversions. 

For the normalization channel $K_{3\pi}^{+-}$ the $M_{3\pi}$ and $p_t$ were inside a smaller 
ellipse with semi-axes $12\ MeV/c^2$ and $25\ MeV/c$, respectively.
A sample of about 1.11 million $K_{e4}^{+-}$ candidates and about 19 millions of prescaled 
$K_{3\pi}$ candidates were selected from data recorded in 2003-2004.

Two main background sources for this mode are known: $K^{\pm} \rightarrow \pi^+ \pi^- \pi^\pm$
decays with subsequent $\pi \rightarrow e\nu$ decay or a pion mis-identified
as an electron; and $K^{\pm} \rightarrow \pi^0 (\pi^0) \pi^\pm$ with subsequent 
$\pi^0 \rightarrow e^+e^-\gamma$ decay with undetected photons and an 
electron mis-identified as a pion. Their admixture in the signal events is estimated to be below 1\%.  

A detailed GEANT3-based \cite{Brun:1994aa} Monte Carlo simulation was used to take into account full 
detector geometry, DCH alignment, local inefficiencies and beam properties. 

The resulting $K_{e4}^{+-}$ branching fraction \cite{Batley:2012rf}
$BR(K_{e4}^{+-}) = (4.257 \pm 0.004_{stat} \pm 0.016_{syst} \pm 0.031_{ext}) 10^{-5}$ is 3 times
more precise than available PDG value \cite{Beringer:1900zz} . It has been used to extract 
the common normalization form factor $f_s$ \cite{Batley:2012rf}.

\section{$K^{00}_{e4}$ decay}
\label{k00e4}

The $K_{e4}^{00}$ rate is measured relative to the $K^{\pm} \to \pi^0\pi^0\pi^{\pm}$ ($K_{3\pi}^{00}$) 
normalization channel. These two modes are collected using the same trigger and with a similar 
event selections. The separation between them occurs only at a later stage.

Events with at least four $\gamma$, detected by LKr, and at least one track, reconstructed from 
spectrometer data, were regarded as $K_{e4}^{00}$ or $K_{3\pi}^{00}$ candidates.
Every combination of 4 reconstructed $\gamma$ with energies $E > 3$~GeV was considered as 
a possible pair of $\pi^0$ decays. 
Reconstructed longitudinal positions $Z_1$ and $Z_2$ of both $\pi^0 \to 2\gamma$ decay 
candidates were required to coincide within $500\ cm$, with their average 
position $Z_n = (Z_1+Z_2)/2$ inside the fiducial volume 106 m long. 

Decay longitudinal position $Z_{ch}$, assigned to the track, 
was defined by the closest distance approach between the track and the beam axis. 
Combined vertex, composed 
of four LKr clusters and one charged track with momentum $p > 5 GeV$, was required to have the 
difference $|Z_n - Z_{ch}|$ less than $800\ cm$. 
If several combinations satisfy the vertex criteria, the case of minimum 
$(\frac{Z_1 - Z_2}{\sigma_n})^2+(\frac{Z_n - Z_{ch}}{\sigma_c})^2$ has been chosen, 
where $\sigma_n$ and $\sigma_c$ are the $Z_n$-dependent widths of corresponding 
distributions.

A track was preliminarily identified as $e^\pm$, if it has an associated LKr cluster with $E/p$ between 0.9 and 1.1,
otherwise $\pi^\pm$ was assumed at the first stage. Further suppression of  pions mis-identified as electrons
is obtained by means of discriminant variable which is a linear combination of $E/p$, shower width and 
energy weighted track-to-cluster distance at LKr front face.

\begin{figure}
\begin{center}
\setlength{\unitlength}{1mm}
\resizebox{\columnwidth}{!}{%
\begin{picture}(100.,50.)                
\includegraphics[width=100mm]{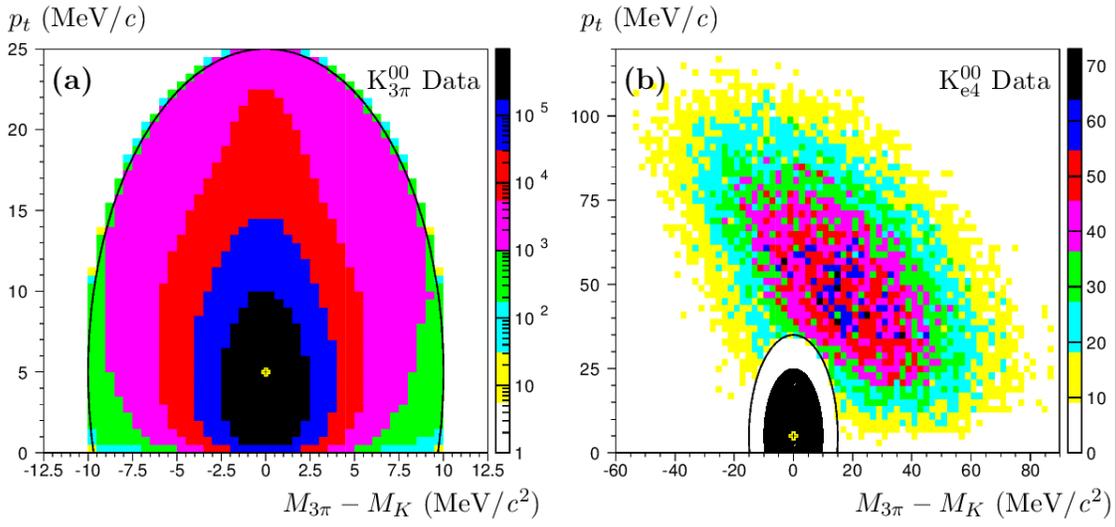}
\end{picture}
}
\end{center}
\caption{
Reconstructed ($M_{3\pi},p_t$) plane for the normalization
(a) and signal (b) candidates. Plot (a) is a zoom inside the smaller ellipse
which defines the normalization sample. Crosses correspond to the ellipse centers
($M_{3\pi}$ = $M_K$; $p_t$ = 5 MeV/$c$).
}
\label{ke4ell}
\end{figure}

$K_{e4}^{00}$ and $K_{3\pi}^{00}$ decays were discriminated by means of elliptic cuts in the 
($M_{\pi^0\pi^0\pi^{\pm}}, p_t$) plane, where $M_{\pi^0\pi^0\pi^{\pm}}$ is the 
invariant mass of combined vertex in the $K_{3\pi}^0$ hypothesis, and $p_t$ is the transversal momentum 
(see Fig. \ref{ke4ell}).
Elliptic cut separates about 94 million $K_{3\pi}^{00}$ normalization events from about
65000 $K_{e4}^{00}$ candidates. Residual fake-electron background is about 0.65\% of $K_{e4}^{00}$
amount. Background from $K_{3\pi}^{00}$ with the subsequent $\pi^{\pm} \to e^{\pm} \nu$ is 
0.12\% of the signal, and the accidental-related background is about 0.23\%. It gives in total 1\% 
of background admixture.

For the case of $K^{\pm} \rightarrow \pi^0 \pi^0 e^{\pm} \nu$ ($K^{00}_{e4}$) decay, 
due to the presence of two identical particles in dipion, it cannot be
in antisymmetric {\it l}=1 state, so form factors do not include P-terms.
In first approximation, only S-wave contributes, and matrix element 
is parametrized in terms of the only formfactor $F_s$, that
may depend on $S_{\pi}$ and $S_e$. Form factor $F_s$ was extracted from the fit of events 
distribution on ( $S_e , S_{\pi}$) plane, taking into account the acceptance, 
calculated from MC simulation.

The following empirical parameterization has been used:
\begin{eqnarray}
\nonumber
F_s/f_s = 1+(f'_s/f_s) q^2 + (f''_s/f_s) q^4 + (f'_e/f_s) S_e/4m^2_{\pi}\ for\ q^2 > 0; \\
F_s/f_s = 1+ d \sqrt{|q^2/(1+q^2)|}   + (f'_e/f_s) S_e/4m^2_{\pi}\ for\ q^2 < 0.
\end{eqnarray}

The results are in agreement with NA48/2 $K_{e4}^{+-}$ analysis described above: 
\begin{eqnarray}
\nonumber
f'_s/f_s = 0.149 \pm 0.033_{stat} \pm 0.014_{syst}, \\ 
\nonumber
f''_s/f_s = -0.070 \pm 0.039_{stat} \pm 0.013_{syst}, \\
\nonumber
f'_e/f_s = 0.113 \pm 0.022_{stat} \pm 0.007_{syst}, \\
\nonumber
d = -0.256 \pm 0.049 \pm 0.016_{syst}.
\end{eqnarray}

\begin{figure}
\begin{center}
\setlength{\unitlength}{1mm}
\resizebox{0.80\columnwidth}{!}{%
\begin{picture}(100.,50.)                
\includegraphics[width=100mm]{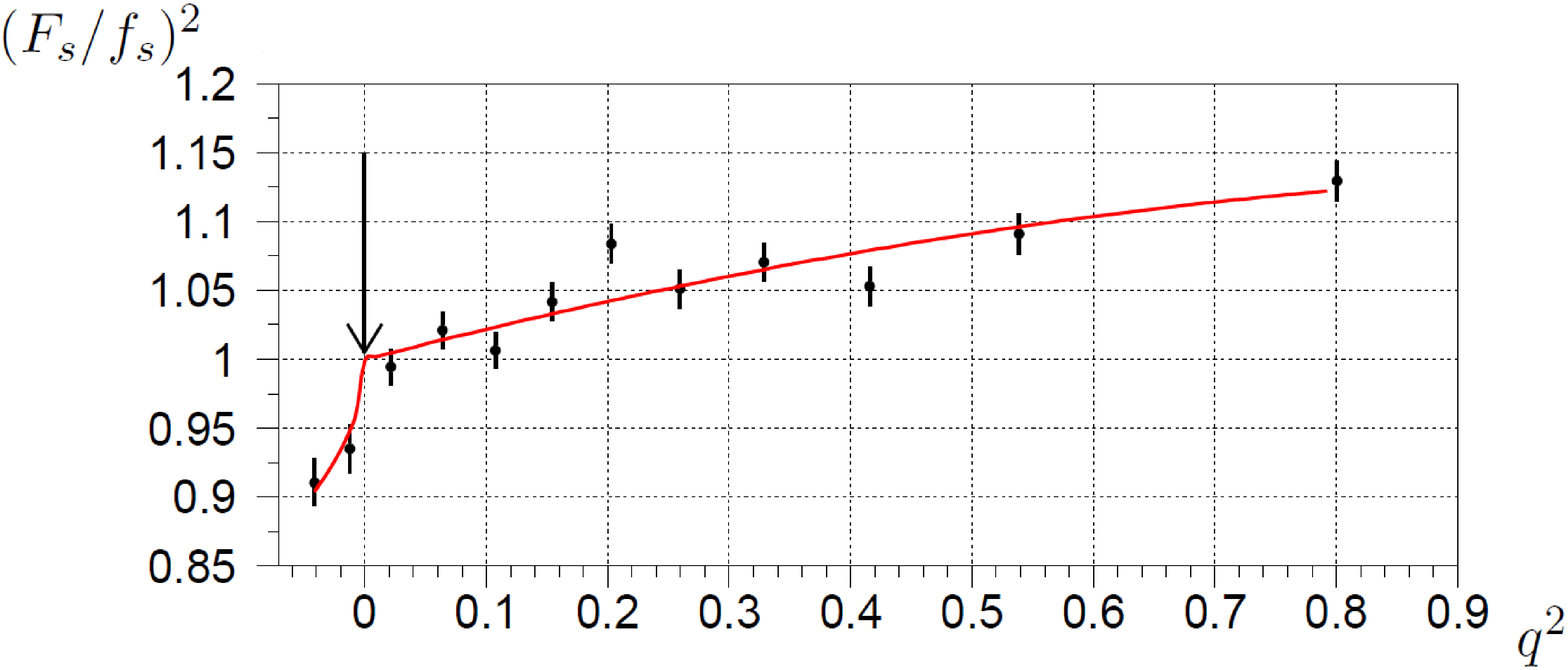}
\end{picture}
}
\end{center}
\caption{ $K_{e4}^{00}$ normalized form factor squared as a function of $q^2$. 
The line corresponds to the adopted empirical fit. The arrow points to the $2 m_{\pi}$ threshold.}
\label{fss}
\end{figure}

The obtained form factor was used to obtain the final result of branching fraction measurement: 
$Br(K_{e4}^{00})=(2.552 \pm 0.010_{stat} \pm 0.010_{syst} \pm 0.032_{ext})10^{-5}$.
It is 10 times more precise, than PDG corresponding value \cite{Beringer:1900zz}. 
Systematic error includes the contributions from background, simulation statistical error, sensitivity to 
form factor, radiation correction, trigger efficiency and beam geometry.
External error comes from uncertainty of normalization channel $K_{3\pi}^{00}$ branching fraction.

Below the threshold of $S_\pi = (2m_{\pi^\pm})^2$ the measured $K_{e4}^{00}$ decay form factor shows
a deficit of events, that is well described by the present empirical parameterization (Fig. \ref{fss}). It is similar 
to the effect of $\pi^+\pi^- \to \pi^0\pi^0$ rescattering in $K^\pm \to \pi^0\pi^0\pi^\pm$ decay 
(cusp effect \cite{Cabibbo:2004gq}), investigated by NA48/2 collaboration earlier \cite{Batley:2009zz} 
on the basis of ChPT formulations.

\section{$K^\pm\to\pi^\pm\gamma\gamma$ decay}

In the ChPT framework, the $K^\pm\to\pi^\pm\gamma\gamma$ decay receives two 
non-interfering contributions at lowest non-trivial order ${\cal O}(p^4)$: the pion and kaon {\it loop amplitude} depending on 
an unknown ${\cal O}(1)$ constant $\hat{c}$ representing the total contribution of the counterterms, 
and the {\it pole amplitude}~\cite{ec88}. 

\begin{figure}
\begin{center}
\setlength{\unitlength}{1mm}
\resizebox{\columnwidth}{!}{%
\begin{picture}(100.,50.)                
\includegraphics[width=100mm]{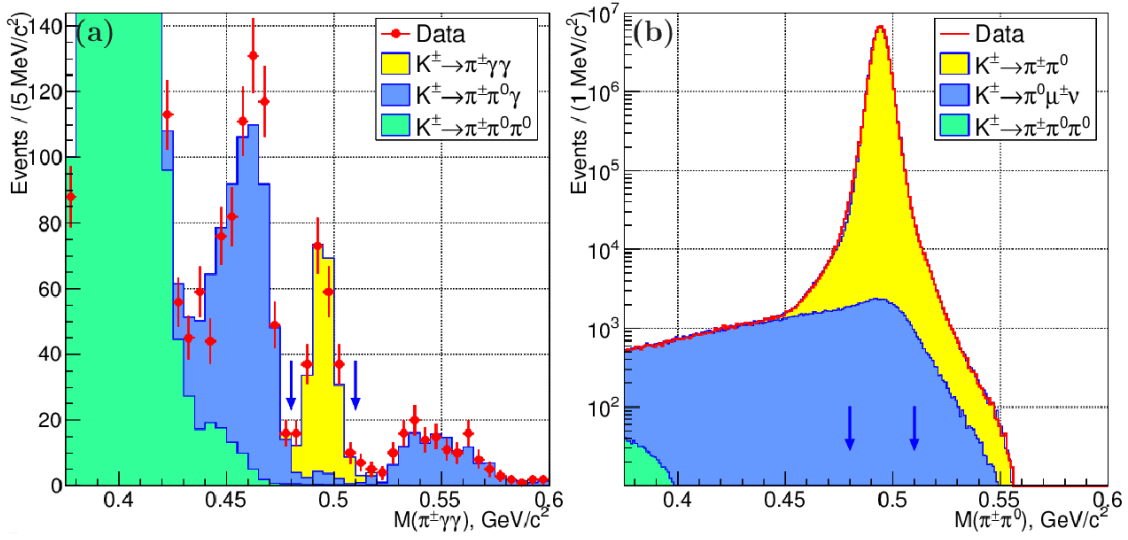}
\end{picture}
}
\end{center}
\caption{Reconstructed invariant mass distributions of $\pi^\pm \gamma \gamma$ (a) and 
$\pi^\pm \pi^0$ (normalization channel) candidates. Filled circles: NA62 ($R_K$ stage) 
experimental data; histograms: simulated signal and background components.
Signal regions are indicated with arrows.}
\label{mpiggnew}
\end{figure}

New measurements of this decay have been performed using data collected during a 3-day special NA48/2 
run in 2004 and a 3-month NA62 run in 2007.
The $K_{\pi \gamma \gamma}$ decay rate has been measured with respect to the normalization decay chain: 
$K^\pm \to \pi^\pm \pi^0$ decay followed by $\pi^0 \to \gamma \gamma$ (Fig. \ref{mpiggnew}). 

\begin{figure}
\begin{center}
\setlength{\unitlength}{1mm}
\resizebox{0.5\columnwidth}{!}{%
\begin{picture}(100.,100.)                
\includegraphics[width=100mm]{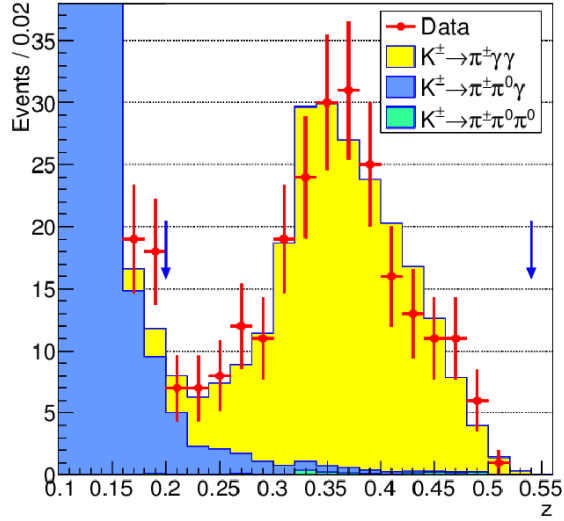}
\end{picture}
}
\end{center}
\caption{Reconstructed $z=(m_{\gamma \gamma} / m_K)^2$ spectrum for selected
$K_{\pi\gamma\gamma}$ candidates compared to the simulated signal and
background distributions. Signal region is indicsted with vertical arrows.}
\label{zdist}
\end{figure}

Both for signal and normalization modes only one reconstructed charged particle track with the
closest distance of approach (CDA) to beam axis less than 3.5 cm and with the momentum $p$ between 
8 and 50 GeV/$c$ was required. The ratio of corresponding LKr cluster energy to the track momentum,
measured by means of spectrometer was $E/p < 0.8$. Two LKr clusters with energies $E>3$~GeV/$c$ in 
time with the track ($\pm 15\ ns$), but separated by at least 25 cm from the track impact point on 
LKr front plane were considered as $\gamma$ candidates. An energy-dependent upper limit was imposed
on the cluster lateral width to suppress the contribution of cases with cluster merging.
 
Signal events are selected in the region of $z=(m_{\gamma\gamma}/m_K)^2>0.2$ to reject the $K^\pm\to\pi^\pm\pi^0$ 
background peaking at $z=0.075$ (see Fig. \ref{zdist}). For $K_{2\pi}$ as a normalization channel, 
$0.064 < z < 0.086$ was required.
149 (232) decays candidates are observed in the 2004 (2007) data set, with backgrounds 
contaminations of 10.4\% (7.5\%) from $K^\pm\to\pi^\pm\pi^0(\pi^0)(\gamma)$ decays with merged photon 
clusters in the electromagnetic calorimeter.

The values of $\hat{c}$ in the frameworks of the ChPT ${\cal O}(p^4)$ and ${\cal O}(p^6)$ 
parameterizations \cite{da96} as well as branching ratio have been measured using likelihood fits to the data. 
The main systematic effect is due to the background uncertainty. Uncertainties 
related to trigger, particle identification, acceptance and accidental effects found to be negligible.
The final combined results based on 2004 and 2007 runs data \cite{pigg1,pigg2} are:
$\hat{c}$ for ${\cal O}(p^4)$ fit = $1.72\pm 0.20_{stat} \pm 0.06_{syst}$;
$\hat{c}$ for ${\cal O}(p^6)$ fit =  $1.86 \pm 0.23_{stat} \pm 0.11_{syst}$; 
branching fraction $Br(K_{\pi\gamma\gamma})$ for ${\cal O}(p^6)$ fit = $(1.003\pm0.056)\times 10^{-6}$. 
The model-independent branching ratio for $z > 0.2$ is equal to $(0.965\pm0.063)\times 10^{-6}$.
New results are in agreement with the earlier (based on 31 events) BNL E787 \cite{bnle787} ones.

\end{document}